\documentclass[a4paper,10pt]{article}
\usepackage[T2A]{fontenc}
\usepackage{amssymb}
\usepackage{amsfonts}
\usepackage{amsmath}
\usepackage{pstricks}
\usepackage{graphicx}
\usepackage{multicol}
\numberwithin{equation}{section}
 \DeclareMathOperator{\sgn}{\rm sgn}
 \DeclareMathOperator{\diag}{\rm diag}
 \renewcommand{\Re}{\text{Re}}\renewcommand{\Im}{\text{Im}}

\title{Induced dynamics}
\author{A.~K.~Pogrebkov${}^{*}$\\
${}^{*}$Steklov Mathematical Institute, \\
and HSE University, Russian Federation,\\
Moscowand Ruijsenaars--Schneider models, cheshirization}

\begin{document}

\maketitle

\begin{abstract}
Induced dynamics is defined as dynamics of real zeros with respect to $x$ of equation 
$f(q_1-x,\ldots,q_N-x,p_1,\ldots,p_N)=0$, where $f$ is a function, and $q_i$ and $p_j$ are canonical variables obeying some 
(free) evolution. Identifying zero level lines with the world lines of particles, we show that the resulting dynamical 
system 
demonstrates highly nontrivial collisions of particles. In particular, induced dynamical systems can describe such 
``quantum'' effects as bound states and creation/annihilation of particles, both in nonrelativistic and relativistic cases. 
On the other side, induced dynamical systems inherit properties of the $(p,q)$-systems being Hamiltonian and Liouville 
integrable. 
\end{abstract}

\section{Introduction}

Let $\mathcal{A}_N$ denote a phase space of $N$-particle dynamical system with coordinates $q_i$ and momenta $p_i$, 
$i=1,\ldots,N$, canonical with respect to the Poisson bracket $\{q_i,p_j\}=\delta_{ij}$. Let $H=H(\mathbf{q},\mathbf{p})$ 
denote the  
Hamiltonian of this system, where $\mathbf{q}=(q_1,\ldots,q_N)$, $\mathbf{p}=(p_1,\ldots,p_N)$, i.e., $\dot 
q_i=\{q_i,H\}$, $\dot p_i=\{p_i,H\}$. We assume that $q_i$ are either real or pairwise complex conjugate and the same are 
properties of the corresponding $p_i$. Let we have a function $f(\mathbf{q},\mathbf{p})$ on  $\mathcal{A}_N$ and let 
$f(\mathbf{q}-x\mathbf{e},\mathbf{p})$ denote this function with all coordinates $q_i$ shifted by a real parameter $x$, 
$\mathbf{e}=(\underbrace{1,\ldots,1}_{N})$. In what follows we assume that function $f$ is such that equation
\begin{equation}
 f(\mathbf{q}-x\mathbf{e},\mathbf{p})=0\label{i2}
\end{equation}
has $M$ simple real zeros $x_1,\ldots,x_M$, where $M\leq N$. We assume also that there exists such open subset 
$\mathcal{A}_{N}'\subset\mathcal{A}_N$, that $M=N$ for any $(\mathbf{q},\mathbf{p})\in\mathcal{A}_{N}'$. We define the 
\textbf{induced system} as 
system with configuration space given by \textbf{real} zeros of (\ref{i2}). This system is dynamical as due to (\ref{i2}) 
all roots $x_i(t)$ are functions on $\mathcal{A}_N$ and depend on $t$ via $\mathbf{q}$ and $\mathbf{p}$ only. Evolution of 
this system is given by 
the same Hamiltonian $H$, $\dot{x_i}=\{x_i,H\}$,  under the same Poisson bracket $\{q_i,p_j\}=\delta_{ij}$. Here, for 
simplicity, we consider the case of a trivial dynamics on $\mathcal{A}$:
\begin{equation}
 H=\sum_{i=1}^{N}h(p_i),\label{i1}
\end{equation}
where $h$ is a function of one variable, so that
\begin{equation}
\dot q_i=h'(p_i),\qquad \dot p_i=0.\label{i3}
\end{equation}
In this case the induced system is not only Hamiltonian but also (super)integ\-ra\-ble, as by construction it has (at least) 
$N$ integrals of motion in involution.

Assume, that $(\mathbf{q},\mathbf{p})\in\mathcal{A}_{N}'$, i.e., there exists exactly $N$ real (different) solutions of the  
Eq.~(\ref{i2}):
\begin{equation}
 f(\mathbf{q}-x_i\mathbf{e},\mathbf{p})=0,\qquad i=1,\ldots,N.\label{i4}
\end{equation}
Taking (\ref{i3}) into account we differentiate (\ref{i4}) twice with respect to $t$:
\begin{align}
&\sum_{j=1}^{N}(h'(p_j)-\dot{x}_{i}) f_{q_j}(\mathbf{q}-x_i\mathbf{e},\mathbf{p})=0,\label{i5}\\
&\ddot{x}_{i}\sum_{j=1}^{N} f_{q_j}(\mathbf{q}-x_i\mathbf{e},\mathbf{p})=
\sum_{j,k=1}^{N}(h'(p_j)-\dot{x}_{i})(h'(p_k)-\dot{x}_{k}) f_{q_jq_k}(\mathbf{q}-x_i\mathbf{e},\mathbf{p}).\label{i6}
\end{align}
One can consider (\ref{i4}) and (\ref{i5}) as system of $2N$ equations on $2N$ unknowns $\mathbf{q}$ and $\mathbf{p}$, that 
are defined by means of these equations as functions of $\mathbf{x}$ and $\mathbf{\dot{x}}$ under condition of unique 
solvability of this system, that we assume below. Inserting these functions in (\ref{i6}) we prove existence of the 
Newton-type equations of \textbf{the induced dynamical 
system}:
\begin{equation}
\ddot{x}_{i}=F_{i}(x_1,\ldots,x_N,\dot{x}_1,\ldots,\dot{x}_N),\quad i=1,\ldots,N,\label{i7}
\end{equation}
where $F_i$ are some forces depending on differences $x_i-x_j$ and, generically, on the velocities $\dot{x}_i$.

The above consideration gives also scheme of solution of the Cauchy problem for the induced system. Let we are given with 
$2N$ initial data: $x_i(0)$ and $\dot{x}_j(0)$, say at $t=0$, where $i,j=1,\ldots,N$. Equations (\ref{i4}) and (\ref{i5}) 
define values $(\mathbf{q}(0),\mathbf{p}(0))$ that belong to $\mathcal{A}_N'$ by definition. Then by (\ref{i3}) 
$q_i(t)=q_i(0)+th'(p_i)$, $p_i(t)=p_i$, that after substitution in (\ref{i2}) gives $M$ real roots $x_1(t),\ldots,x_M(t)$ 
for 
any $t\in\mathbb{R}$. Notice that $M$ is not obliged to be equal to $N$ at any moment of time, i.e., point 
$(\mathbf{q}(t),\mathbf{p}(t))$ is not obliged to belong to $\mathcal{A}_N'$. Thus the scheme of solution of the Cauchy 
problem 
for the induced system is close to the one for integrable nonlinear PDE's.

Below we present examples that demonstrate that in spite of a trivial dynamics of the system on the phase space 
$\mathcal{A}$, 
dynamics of the induced system is highly nontrivial. In particular, it demonstrates such ``quantum'' effects as existence of 
 
stable bound states, processes of creation and annihilation of particles and some other. We show that generically these 
dynamical systems are systems with variable dimension of their configuration space. In this sense the original space 
$\mathcal{A}_N$ 
plays the role of a moduli space of the induced dynamical system. The manuscript is organized as follows. In Sec.~2 we 
consider dynamical systems that appeared many years ago (see, e.g., \cite{kdv},\cite{sinh1},\cite{sinh2},\cite{sinh3}) as 
systems describing the dynamics of singularities of solutions of some integrable equations (``zeros'' of $\tau$-functions). 
More exactly, we consider here examples of the KdV  and Sinh--Gordon equations and show that dynamics of their singular 
soliton solutions give examples of the induced systems. In Sec.~3 we show that the same is valid for the famous dynamical 
systems: rational Calogero--Moser, \cite{CM1,CM2}, and Ruijsenaars--Schneider, \cite{RS1,RS2}, models. In Sec.~4 we present 
some other simple examples of the induced systems given by polynomial functions $f$ in (\ref{i2}), both in nonrelativistic 
and relativistic cases. Possible generalizations of the suggested approach are discussed in Sec.~5. Properties of the 
induced systems under consideration are displayed by means of figures carried out by package Wolfram Mathematica 11.1.

\section{Dynamics of singularities of integrable differential equations.}
\subsection{Singular solutions of the KdV equation.}
In this section we we briefly present old results on dynamics of singularities of soliton solutions (singular solitons) of 
two integrable PDEs. We start with $N$-soliton solution of the KdV equation 
\begin{equation}
4u_{t}-6uu_{x}+u_{xxx}=0,\label{a13}
\end{equation}
on real function $u(t,x)$, where indexes denote partial derivatives. This famous equation is known to have regular 
soliton solutions (see, e.g., \cite{nov}) and singular ones, \cite{kdv}. Its generic $N$-soliton solution is given by
\begin{equation}
 u(t,x)=-\partial_{x}^{2}\log\det(E(t,x)+v)^{2},\qquad \label{a14}
\end{equation}
where  $E(t,x)$ and $v$ are, correspondingly, $N\times{N}$ diagonal and constant matrices 
\begin{equation}
E(t,x)=\diag\bigl\{\epsilon_{i}e^{2{p}_{i}(x-a_i-{p}_{i}^{2}t)}\bigr\},\qquad 
v_{ij}=\dfrac{2{p}_{i}}{{p}_{i}+{p}_{j}}, \label{a15}
\end{equation}
where $a_i$, ${p}_i$, and $\epsilon_i=\pm1$, $i=1,\ldots,N$ are constant parameters of the solution such that 
$\Re{p}_{i}>0$ and
\begin{equation}\label{a16}
\begin{split}
 &\text{either } \Im{p}_{i}=0, \quad\epsilon_i=\pm1,\qquad \Im a_i=0,\\
 &\text{or } \Im{p}_{i}\neq0,\text{ then there exists } {p}_{l}=\overline{p}_{i}, \quad\epsilon_l=\epsilon_i=\pm1,\quad 
a_l=\overline{a}_i.
\end{split}
\end{equation}
In the case where $\Im{p}_{i}=0$ signs $\epsilon_i=+1$ and $\epsilon_i=-1$ correspond to the regular and singular solitons.  
Every pair of ${p}_i=\overline{p}_l$ with $\Im{p}_{i}\neq0$ gives one line of singularity. All singularities of the solution 
$u(t,x)$ are given by the zeros of determinant in (\ref{a14}),
\begin{equation}
 \det(E(t,x)+v)=0.\label{a17}
\end{equation}
Thus in generic situation relations (\ref{a14}), (\ref{a15}) describe interaction of regular and singular solitons and 
breathers, corresponding to mutually conjugate pairs of parameters. In \cite{kdv} it was shown that singularities of this 
solution form smooth curves on $(x,t)$-plane. They run from minus to plus $t$-infinities and thus are observable at any 
moment of time. On the other side, it is well known that regular solitons are observable outside the collision region only. 
Taking that asymptotic behavior of the regular and singular solitons coincides into account, we suggested to introduce 
``charge conjugation'' (see~\cite{kdv}), i.e., to change $E(t,x)+v\to E(t,x)-v$. This substitution is equivalent to change 
of 
all signs $\epsilon_i\to-\epsilon_i$, so regular and singular solitons mutually exchange and we get world lines of particles 
corresponding to both, regular and singular solitons, as zeros of the product
\begin{equation}
\det(E(t,x)+v)\det(E(t,x)-v)=0,\label{a18}
\end{equation}
instead of (\ref{a17}). Omitting the trivial case of one-soliton solution that gives free motion, we present evolutions of 
two and three singularities, i.e. systems of two and three particles. 
\begin{figure}[t]
\begin{multicols}{2}
\hfill
\includegraphics[width=50mm]{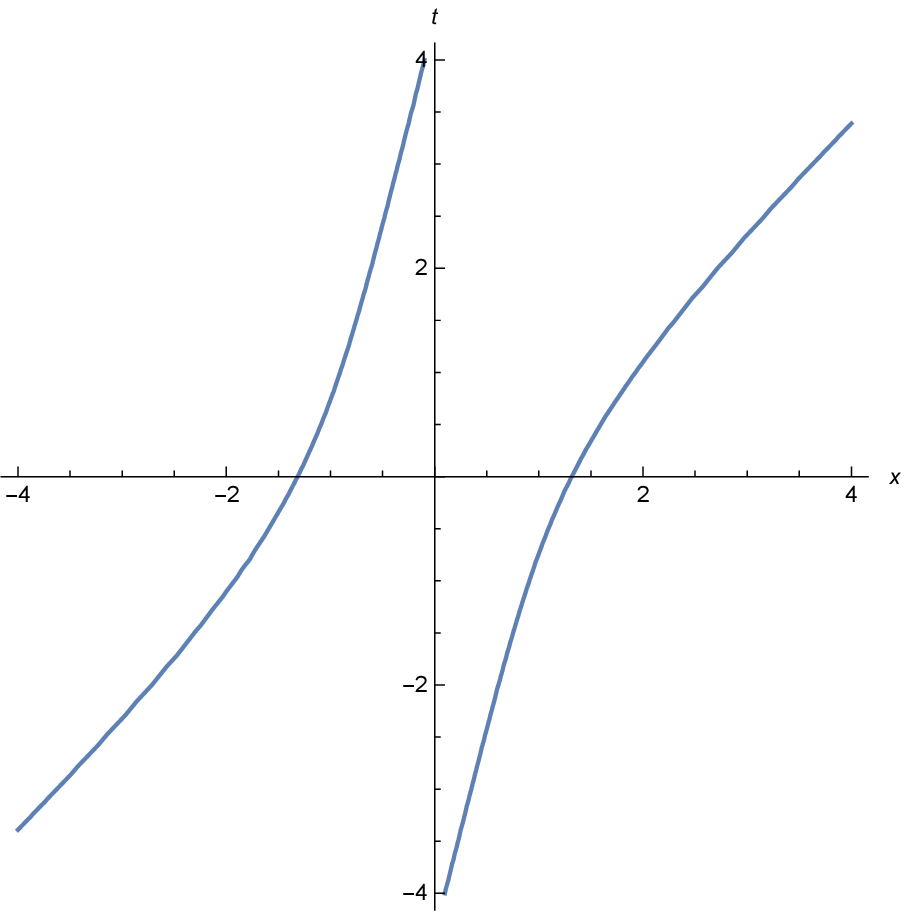}
\hfill
\caption{Soliton--soliton collision, $\epsilon_1=\epsilon_2=1$.}
\label{ss}
\hfill
\includegraphics[width=50mm]{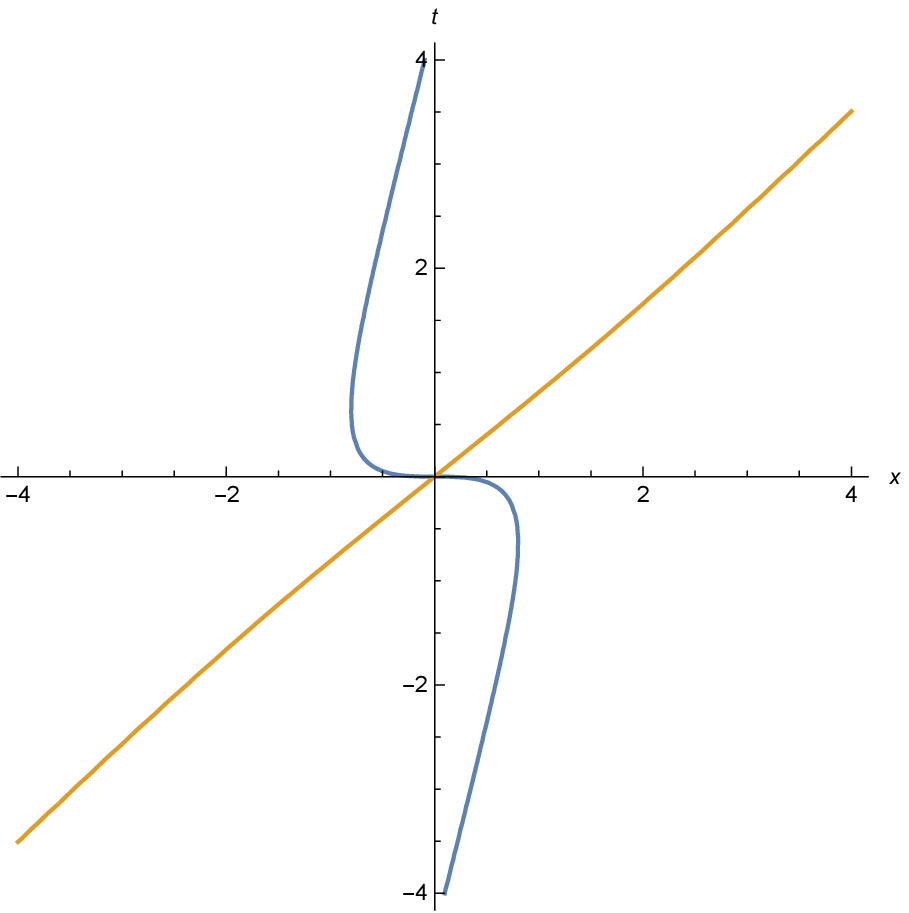}
\hfill
\caption{Soliton--antisoliton collision: $\epsilon_1=-\epsilon_2$}
\label{sa}
\end{multicols}
\end{figure}
Thus on Fig.~\ref{ss} we have soliton--soliton collision. We see that solitons move with a finite velocity, 
stop and repulse, taking at infinity phase shifts. These two curves are zeros of one of the factors in (\ref{a18}), 
while the second factor has no zeros in this case. Fig.~\ref{sa} presents collision of regular soliton with the singular 
one (soliton--antisoliton collision), i.e., the case were every factor in (\ref{a18}) has just one zero at any value of $t$. 
These regular and singular solitons attract, their world lines intersect. In a vicinity of intersection the slowest soliton 
moves to the left and at the point of intersection its velocity is infinite. 
\begin{figure}[ht]
\begin{multicols}{2}
\hfill
\includegraphics[width=50mm]{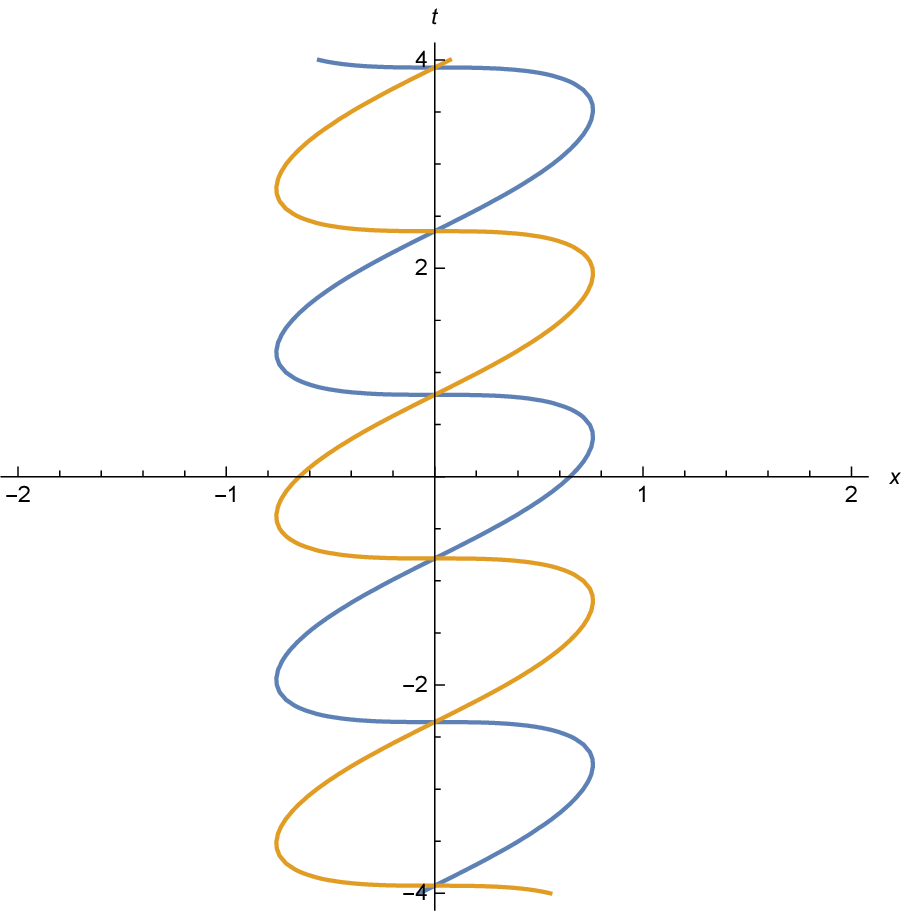}
\hfill
\caption{KdV: breather}
\label{br}
\hfill
\includegraphics[width=50mm]{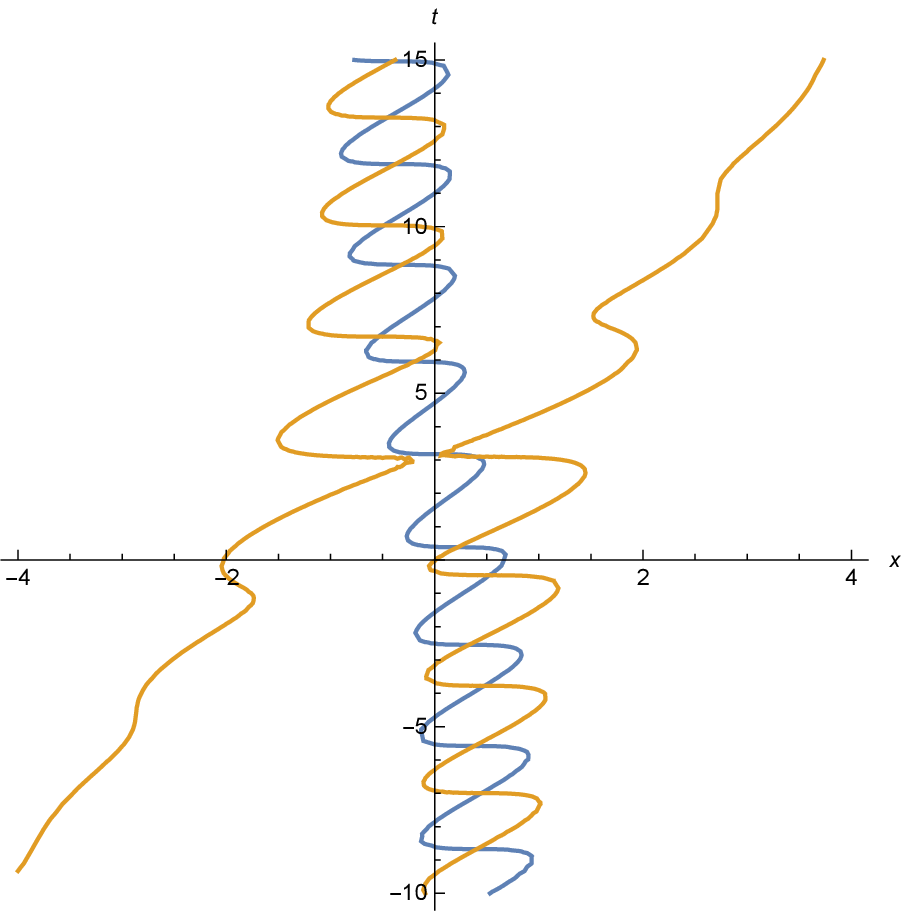}
\hfill
\caption{Kdv: soliton--breather collision}
\label{sbr}
\end{multicols}
\end{figure}

World lines of singularities of the breather solution are presented on the Fig.~\ref{br}. Again, we have two curves here, 
every curve appears as set of zeros of one of multiples in the l.h.s.\ of (\ref{a18}). The particles attract and circulate 
around the common center passing one through another. The particle moving to the left has infinite velocity at the point of 
intersection. Fig.~\ref{br} shows also that in contrast to the previous cases one can choose for the breather a 
center-of-mass frame. Finally, on Fig.~\ref{sbr} we present interaction of (anti)soliton and breather. This interaction is 
highly nontrivial: the swooping soliton knocks out soliton of the same nature (given by the zero of the same factor in 
(\ref{a18})) and fuse in a bound state with the second partner of the original breather. At infinity both breather and 
soliton get phase shifts. 

Let us introduce functions
\begin{equation}\label{a19}
q_{i}(t)=a_{i}+p_{i}^{2}t,\qquad i=1,\ldots,N, 
\end{equation}
that enables us to rewrite (\ref{a18}) in the form (\ref{i2}), where
\begin{equation}
 f(\mathbf{q},\mathbf{p})=\det(E_0+v) \det(E_0-v),\qquad E_0=\diag\bigl\{\epsilon_{i}e^{-2{p}_{i}q_i(t)}\bigr\},\label{a20}
\end{equation}
cf.\ (\ref{a15}). We see that in analogy to Introduction this dynamical system is Hamiltonian, i.e., obeys (\ref{i1}) with 
respect to the canonical Poisson bracket $\{q_i,{p}_j\}=\delta_{ij}$. Its Hamiltonian (cf.\ (\ref{i1})) equals
\begin{equation}\label{a21}
H=\dfrac{1}{3}\sum_{i=1}^{N}{p}_{i}^{3},
\end{equation}
and the system is Liouville integrable: variables $\{p_1,\ldots,p_N\}$ are integrals and they are in involution by 
construction. It is easy to prove that system (\ref{i5}), (\ref{i6}) for the function $f$ in (\ref{a20}) is solvable with 
respect to $q_i$ and $p_i$, while not explicitly. Correspondingly, here forces (i.e., the r.h.s.\ of (\ref{i7})) exist but 
in implicit form. It is Eq.~(\ref{i2}) that gives the most descriptive characterization of the induced dynamical system. 

\subsection{Singular solutions of the Sinh--Gordon equation}
The same consideration is applicable in the relativistic case. Here we consider a relativistic induced dynamical system 
given by motion of singularities of the soliton solutions of the Sinh--Gordon equation, $u_{tt}-u_{xx}+\frac{1}{2}\sinh 
u=0$, on the real function $u(t,x)$. Its  $N$-soliton solution, \cite{sinh2} and \cite{sinh3}, is given by
\begin{equation}\label{a2}
\begin{split}
&e^{u(t,x)}=\dfrac{\det(E(\xi,\eta)+v)}{\det(E(\xi,\eta)-v)},\\
& E(\xi,\eta)=\diag\bigl\{\epsilon_{i}e^{2[(\xi-a_i){p}_i+\eta/{p}_i]}\bigr\}_{i=1}^{N},
\end{split}
\end{equation}
where 
\begin{equation}
\xi=x+t,\qquad \eta=x-t,\label{a2'}
\end{equation}
are cone variables, matrix $v$ is given in (\ref{a15}) and parameters in the r.h.s.\ obey conditions (\ref{a16}). In this 
case regular solitons do not exist, in contrast to the KdV case. Instead, we have here singularities given by zeros of both 
determinants in (\ref{a2}), so these singularities are given by zeros of the product (cf.\ (\ref{a18}))
\begin{equation}
\det(E(\xi,\eta)+v)\det(E(\xi,\eta)-v)=0.\label{a3}
\end{equation}
These zeros form $N$ smooth time-like curves $\xi_i(\eta)$, $i=1,\ldots,N$. Zeros of the first and second factors give 
$u=-\infty$ and $u=+\infty$ correspondingly. Lines of singularities of the different signs can intersect and at these 
points and only at these points they are light-like. Real parameters ${p}_i$ correspond to solitons, or anti-solitons 
(depending on the signs $\epsilon_i$). Mutually conjugate pairs of these parameters corresponds to breathers. 

Setting
\begin{equation}
 q_{i}(\eta)=a_{i}-\dfrac{\eta}{{p}^{2}_i},\label{a4}
\end{equation}
we can write matrix $E(\xi,\eta)$ in (\ref{a2}) in the form 
$E(\xi,\eta)=\diag\bigl\{\epsilon_{i}e^{2{p}_i(\xi-q_i)}\bigr\}_{i=1}^{N}$. Thus equation (\ref{a3}) has the form of 
(\ref{i2}),
\begin{equation}
f(\mathbf{q}-\xi\mathbf{e},\mathbf{p})=0,\label{a8} 
\end{equation}
where
\begin{equation}
f(\mathbf{q},\mathbf{p})=\det(E_0+v)\det(E_0-v),\qquad E_0=\diag\bigl\{\epsilon_{i}e^{-2{p}_iq_i}\bigr\}_{i=1}^{N}.\label{a5}
\end{equation}
This function coincides with the one given in (\ref{a20}) up to substitution $x\to\xi$, $t\to\eta$ but dynamics on the space 
$\mathcal{A}$ is given here by
\begin{equation}
 q_i'(\eta)=-\dfrac{1}{{p}_i^2},\qquad {p}_i'=0,\quad i=1,\ldots,N,\label{a6}
\end{equation}
instead of (\ref{a19}). The induced system is Hamiltonian, $H=\sum_{i=1}^{N}{p}_i^{-1}$ (cf.\ (\ref{i1})), 
with respect to the canonical Poisson bracket $\{q_i,p_j\}=\delta_{ij}$. Time evolution of the system of zeros of (\ref{a8}) 
is highly nontrivial: particles repulse, attract, form bound states (breathers) and get nontrivial phase shifts at infinity. 

Cone variables $\xi$ and $\eta$, (\ref{a2'}), enable explicit control of the relativistic invariance of the induced 
system (\ref{a8}). In these terms Lorentz boost is given by rescaling: 
\begin{equation}\label{a7}
\xi\to\lambda\xi,\quad\eta\to\lambda^{-1}\eta,\quad q_i\to\lambda q_i,\quad p_i\to\lambda^{-1}p_i, 
\end{equation}
where $\lambda$ is an arbitrary positive parameter. It is clear that the description above can be equivalently reformulated 
in terms of the laboratory coordinates $x$ and $t$, where zeros of (\ref{a8}) are given by $N$ curves $x_i(t)$, like in the 
KdV case. We omit figures for two and three particle interaction, as in the $(x,t)$-terms they are very close to the ones on 
Fig.~\ref{ss}--Fig.~\ref{sbr}. The only differences are: (anti)solitons here can move to the right and to the left, their 
velocities obey $|\dot{x}_i|\leq1$, where equality takes place at the points of intersection only. Thus we have interaction 
of massive relativistic particles, see \cite{sinh1}--\cite{sinh3} for details. In particular, we presented there equation of 
motion for the case $N=2$ in the special frame $x_1(t)+x_2(t)=0$:
\begin{equation}
\dfrac{\ddot{x}_{12}\sgn{x_{12}}}{\sqrt{4-{\dot{x}_{12}}^{2}}}=\dfrac{4\varepsilon}{\cosh\biggl(\dfrac{4x_{12}}{\sqrt{4-{
\dot {x}_{12}}^ {2}}}
\sqrt{1+\dfrac{\ddot{x}_{12}\sgn{x_{12}}}{\sqrt{4-{\dot{x}_{12}}^{2}}}}\biggr)-\varepsilon},\label{a10}
\end{equation}
where $x_{12}(t)=x_1(t)-x_2(t)$ and where $\varepsilon=1$ for the case of repulsion and $\varepsilon=-1$ for the both, 
soliton-antisoliton and breather, cases of attraction. In contrast to (\ref{i7}) we have here irrational dependence on 
$\ddot{x}_{12}$, but in spite of this (\ref{a10}) can be easily solved, \cite{sinh3}. 

We considered induced dynamical systems generated by evolution of singularities of the integrable partial differential 
equations. The origin from the theory of integrable PDE's was essential for specific properties of these dynamical systems. 
On the other side in both cases we investigated not the solutions themselves (and not their $\tau$-functions), but products 
of the kind (\ref{a18}), (\ref{a3}). Below we consider other examples of induced dynamical systems, i.e., other 
choices of the function $f$ in (\ref{i2}). 

\section{Rational cases of the Calogero--Moser and Ru\-i\-j\-senaars--Schneider models.}
Let us show that the rational versions of the  famous models of Calogero--Moser (CM) and Ruijsenaars--Schneider (RS) also 
give examples of the induced dynamics, i.e., their solutions are given as roots of Eq.~(\ref{i2}) with proper choice of 
functions $f(\mathbf{q},\mathbf{p})$ and $h$ in (\ref{i2}) and (\ref{i1}). Dynamics of these models, see 
\cite{CM1,CM2,RS1,RS2}, is 
given by equations
\begin{align}
 \text{CR:}\qquad&\ddot{x}_{j}=\sum_{\substack{k=1,\\k\neq j}}^{N}\dfrac{2\gamma^2}{(x_k-x_j)^{3}},\label{cm0}\\
 \text{RS:}\qquad&\ddot{x}_{j}=\sum_{\substack{k=1,\\k\neq j}}^{N}\dfrac{2\gamma^2\dot{x}_{j}\dot{x}_{k}}
{(x_j-x_k)\bigl(\gamma^{2}-(x_j-x_k)^{2}\bigr)}.\label{rs0}
\end{align}
Both systems are completely integrable, see citations above, they have Lax pairs and their $L$-operators can be 
written as
\begin{equation}
 L(t)=\diag\{\dot{x}_1(t),\ldots,\dot{x}_N(t)\}+V(t),\label{cm1}
\end{equation}
where for CM and RS models
\begin{equation}
 V_{\text{CM}}(t)=\biggl(\dfrac{\gamma}{x_{k}(t)-x_{j}(t)}\biggr)_{\substack{j,k=1,\\k\neq j}}^{N},\quad
 V_{\text{RS}}(t)=\biggl(\dfrac{\gamma\dot{x}_k(t)}{x_{k}(t)-x_{j}(t)+\gamma}\biggr)_{\substack{j,k=1,\\k\neq j}}^{N}. 
\label{cmrs1}
\end{equation}
In \cite{OP} and \cite{RS3} it was shown that solutions $x_{i}(t)$ of equations (\ref{cm0}) and (\ref{rs0}) correspondingly 
are eigen values of the matrix $X(0)+tL(0)$, where
\begin{equation}
 X(t)=\diag\{x_1(t),\ldots,x_N(t)\},\label{cm2}
\end{equation}
and where $L(0)$ and $L(0)$ are values of these matrices given in terms of the initial data $x_i(0)$, $\dot{x}_i(0)$. In  
\cite{OP} and \cite{RS3} it was proved that, say, at $t\to-\infty$ solutions obey asymptotic behavior 
\begin{equation}
 x_i(t)=a_i+tp_i+O(t^{-1}),\label{cm3}
\end{equation}
where $a_i$ and $p_i$ are constants. 

Let us prove that CM and RS systems can be written in the form (\ref{i2}).  First, notice that due to the translation 
invariance, solutions $x_i(t)$ are roots of the characteristic equation
\begin{equation}
 \det\bigl(X(\tau)+(t-\tau)L(\tau)-xI\bigr)=0,\label{cm4}
\end{equation}
where $\tau$ is an arbitrary initial moment of time. Second, consider limit of (\ref{cm4}) when $\tau\to-\infty$. Thanks to 
(\ref{cm1}) and (\ref{cm3}) we have that
\begin{equation}
 x_i(\tau)+(t-\tau)\dot{x}_i(\tau)=a_i+tp_i+O(\tau^{-1}),\quad \tau\to-\infty.\label{cm5}
\end{equation}
Now for the term $(t-\tau)V(\tau) $ (see (\ref{cm1}) and (\ref{cmrs1})) we have because of (\ref{cm3}) in this limit:
\begin{align}
\dfrac{(t-\tau)\gamma}{x_k(\tau)-x_j(\tau)}&\to \dfrac{\gamma}{p_j-p_k},\label{cm6}\\
\dfrac{(t-\tau)\gamma\dot{x}_j(\tau)}{x_k(\tau)-x_j(\tau)+\gamma}&\to \dfrac{\gamma p_j}{p_j-p_k}.\label{rs1}
\end{align}
Thus solutions of the rational versions of CM and RS models are given by the roots of the equation
\begin{equation}
 \det\bigl(Q+W-xI\bigr)=0,\label{cm7}
\end{equation}
where 
\begin{align}
 &Q(t)=\diag\{q_1(t),\ldots,q_N(t)\},\qquad q_i(t)=a_i+tp_i,\label{cm8}\\
 &W_{\text{CM}}=\biggl(\dfrac{\gamma}{p_{j}-p_{k}}\biggr)_{\substack{j,k=1,\\k\neq j}}^{N},\qquad
 W_{\text{RS}}=\biggl(\dfrac{\gamma p_j}{p_{j}-p_{k}}\biggr)_{\substack{j,k=1,\\k\neq j}}^{N}. \label{cm9}
\end{align}

Characteristic equation (\ref{cm7}) is exactly of the form (\ref{i2}), where $f(\mathbf{q},\mathbf{p})=\det(Q+W)$, 
$\dot{q}_i=p_i$ and $\dot{p}_i=0$, $i=1,\ldots,N$, so that $(q_i,p_j)$ are canonical variables with respect to the bracket 
$\{q_i,p_j\}=\delta_{i,j}$ and Hamiltonian (\ref{i1}) with $h(p)=p^{2}/2$ for both models. Thus we have another proof of the 
Liouville integrability for both these models. Moreover, integrability takes place for any choice of the (off-diagonal) 
matrix $W(\mathbf{p})$ in (\ref{cm7}) that obeys conditions of solvability of the systems (\ref{i4}) and (\ref{i5}). 
Specific 
property of the CM and RS models is possibility to write down explicitly equations (\ref{cm0}) and (\ref{rs0}) of motion 
and Lax pairs, cf.\ discussion of this problem in Sec.~2 for the dynamics of singularities of the KdV and Sinh--Gordon 
equations. 

Here we considered the case of rational versions of CM and RS models only. Derivation for the hyperbolic versions of 
these models can be performed along the same lines. But for the elliptic case construction is more involved, and should be 
based on results of the works \cite{Kr} and \cite{GP}, where solution of these models are given as zeros of a certain 
Riemann theta-function.

\section{Polynomial examples of induced dynamics.}
Here we consider some simplest examples of induced dynamical systems, by polynomial function $f$ in (\ref{i2}), not related 
with solution of any PDE: 
\begin{equation}
 f(\mathbf{q},\mathbf{p})=\prod_{i=1}^{N}q_i-C,\label{p1}
\end{equation}
where $C$ is a real constant (parameter of the system), and variables $q_i$ and $p_i$ obey condition of reality, given in 
the beginning of Sec.~1. Such systems are more trivial then those considered in Sec.\ 2 say, they do not give nontrivial 
phase shifts. Nevertheless, examples of $N=2$ and $N=3$ of the system given by (\ref{p1}) and its relativistic analog (see 
Sec.\ \ref{relat} below) demonstrate such unexpected properties as creation/annihilation of particles. Notice that the phase 
space $\mathcal{A}_N$ here is flat, i.e., there are no conditions of the kind $\Re{p}_i>0$ in (\ref{a16}).

\subsection{Nonrelativistic case.}\label{nonrel}
Let us start with $N=2$ in (\ref{p1}), i.e., function $f$ equal $f(\mathbf{q},\mathbf{p})=q_1q_2-C/4$, that also coincides 
with the $N=2$ case of (\ref{cm7}), where matrix $W$ is $p$-independent. Eq.~(\ref{i2}) in this case sounds as 
\begin{equation}\label{p2}
f(\mathbf{q}-x\mathbf{e},\mathbf{p})\equiv(q_1-x)(q_2-x)-C/4=0.                       
\end{equation}
Let the dynamic system on $\mathcal{A}_2$ be free, given by the Hamiltonian $H=(p_1^{2}+p_2^{2})/2$ (cf.\ (\ref{i1})), so 
that 
$\dot{q}_i=p_i$, $\dot{p}_i=0$. Thus Eq.~(\ref{i2}) is of the second order with respect to $x$, so we have either two real 
solutions $x_1(t)$ and $x_2(t)$, or neither one. Introducing notation for the differences:
\begin{equation}
 x_{12}=x_1-x_2,\qquad q_{12}=q_1-q_2,\label{p3}
\end{equation}
we get
\begin{equation}\label{p4}
x_1+x_2=q_1+q_2,\qquad x_{12}^2=q_{12}^{2}+C,
\end{equation}
so that
\begin{equation}\label{p5}
\dot{x}_1+\dot{x}_2=p_1+p_2,\qquad p_1-p_2=\dfrac{\dot{x}_{12}x_{12}}{q_{12}}. 
\end{equation}
Eqs.~(\ref{p4}) and (\ref{p5}) define $q_i$ and $p_i$ in terms of $x_i$ and $\dot{x}_i$ (cf.\ (\ref{i5}), (\ref{i6})). These 
values, being substituted in the time derivative of (\ref{p5}) gives explicit equations of motion of the induced dynamical 
system, cf.\ (\ref{i7}):
\begin{equation}\label{p6}
\ddot{x}_1+\ddot{x}_2=0,\qquad \ddot{x}_{12}=\dfrac{C\dot{x}^{2}_{12}}{x_{12}(x_{12}^{2}-C)}.
\end{equation}
It is worth to mention that under reduction to the center of mass frame, $\dot{x}_1+\dot{x}_2=0$, these equations coincide 
with the the same reduction of the $N=2$ case of Ruijsenaars--Schneider system, (\ref{rs0}).

It is easy to see that Eqs.~(\ref{p6}) are Lagrangian
\begin{equation}\label{p7}
\mathcal{L}=\dfrac{\dot{x}^2_1+\dot{x}^2_2}{2}+\dfrac{C\dot{x}_{12}^{2}}{4(x_{12}^{2}-C)},
\end{equation}
that in its turn enables to introduce momenta conjugate to $x_i$ as
\begin{equation}
 P_{i}=\dot{x}_{i}+(-1)^{i+1}\dfrac{C\dot{x}_{12}}{2(x_{12}^{2}-C)},\quad i=1,2.\label{p8}
\end{equation}
Thus equations (\ref{p4}) and
\begin{equation*}
p_{1}+p_{2}=P_{1}+P_2,\qquad p_{1}-p_{2}=\dfrac{(P_1-P_2)\sqrt{x_{12}^{2}-C}}{x_{12}}
\end{equation*}
give canonical transformation from variables $\{x_i,P_j\}$ to $\{q_k,p_l\}$, where $i,j,k,l=1,2$. The Hamiltonian $H$, being 
trivial in terms of the variables on $\mathcal{A}_2$ equals
\begin{equation}\label{p9}
 H=\dfrac{P^{2}_{1}+P_2^{2}}{2}-\dfrac{C(P_1-P_2)^{2}}{4x^{2}_{12}},
\end{equation}
in terms of the variables on the phase space of the induced system.
\begin{figure}[ht]
\begin{multicols}{2}
\hfill
\includegraphics[width=50mm]{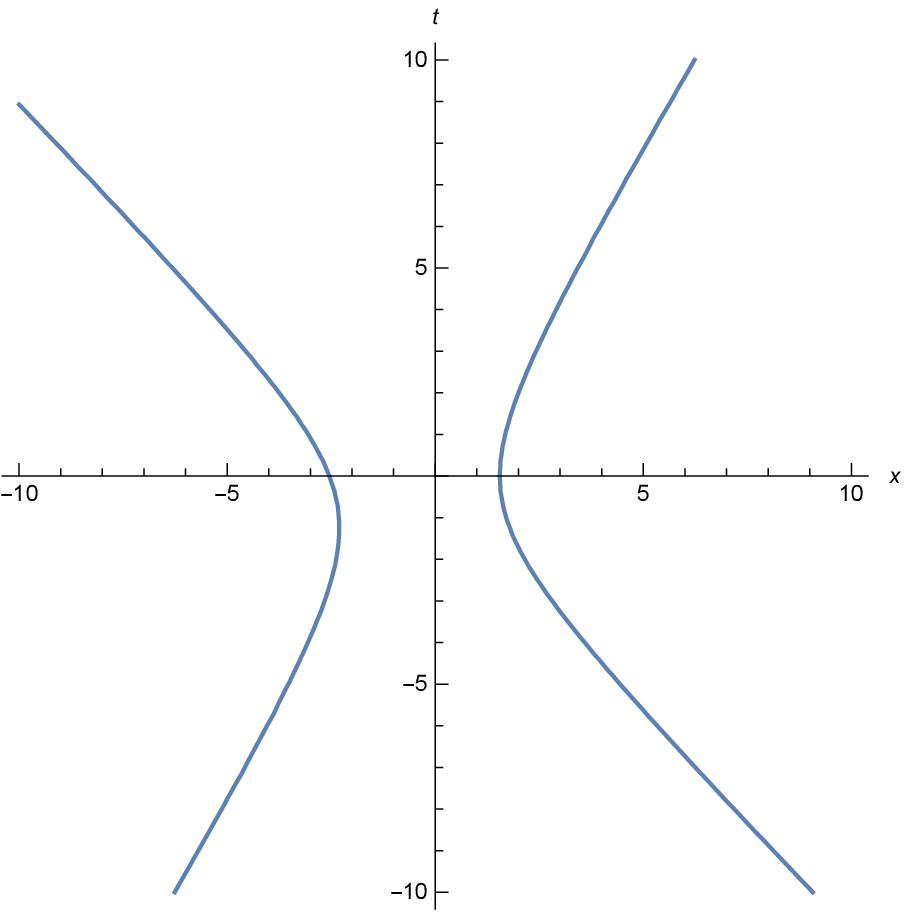}
\hfill
\caption{Two particles, $x_{12}^{2}>C>0$.}
\label{2p>>0}
\hfill
\includegraphics[width=50mm]{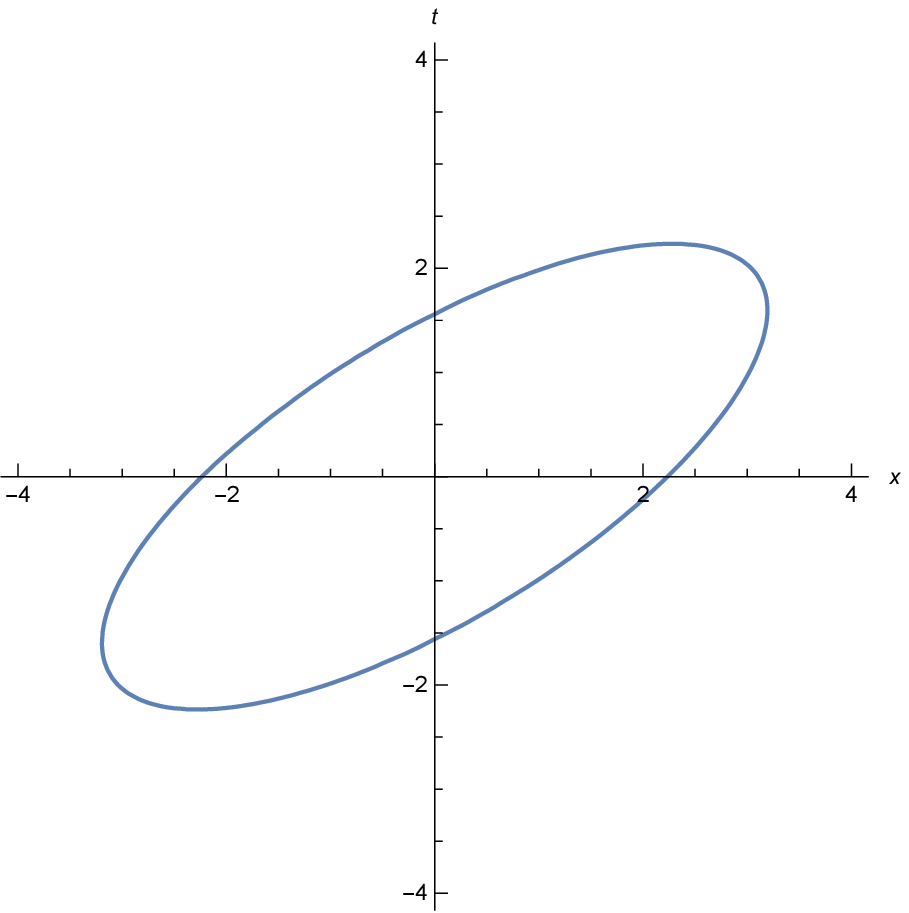}
\hfill
\caption{Two particles, $C>x_{12}^{2}>0$.}
\label{2p>0}
\end{multicols}
\end{figure}

\begin{figure}[ht]
\begin{multicols}{2}
\hfill
\includegraphics[width=50mm]{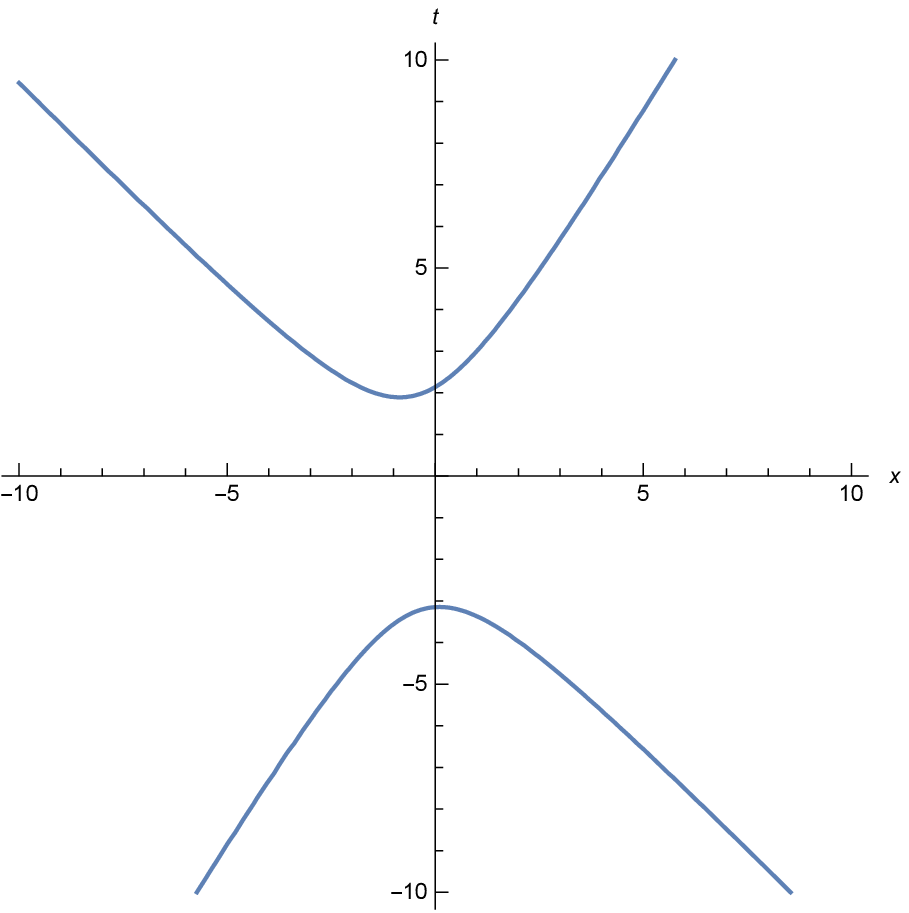}
\hfill
\caption{Two particles, $C<0$}
\label{2p<0}
\hfill
\includegraphics[width=50mm]{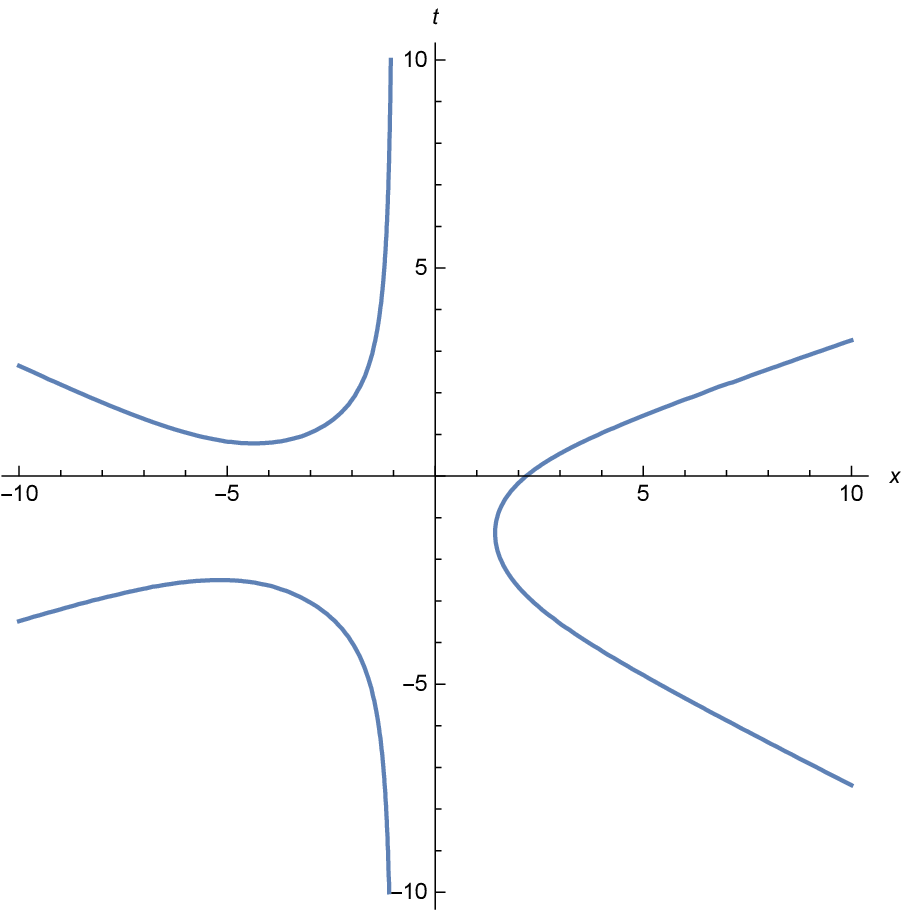}
\hfill
\caption{Three particles.}
\label{3p}
\end{multicols}
\end{figure}

Let us consider initial problem, i.e., we set $x_i(t)|_{t=0}=x_{0,i}$, $\dot{x}_i(t)|_{t=0}=v_{i}$, where $x_{0,i}$ and 
$v_{i}$ are real initial data. Eqs.~(\ref{p4}) and (\ref{p5}) define $q_{i}(0)$ and $p_{i}(0)$, then taking free evolution 
on $\mathcal{A}_2$ into account, we find $q_{i}(t)=q_{i}(0)+p_{i}t$, $p_{i}(t)=p _{i}$, and finally we reconstruct 
$x_{i}(t)$ by (\ref{p4}):
\begin{align}
&x_{1}(t)+x_{2}(t)=x_{0,1}+x_{0,2}+(v_{1}+v_{2})t,\label{p10}\\ 
&x_{12}^{2}(t)=(x_{0,12}+v_{12}t)^{2}+\dfrac{(Cv_{12}t)^{2}}{x_{0,12}^{2}-C},\label{p11}
\end{align}
where we used notation for differences like in (\ref{p3}). Behavior of the roots of Eq.~(\ref{p2}) on the $(x,t)$-plane is 
determined by the sign of $x_{0,12}^{2}-C$. For the induced system with $C>0$ we have in the case of 
$x_{0,12}^{2}>C$ that solutions $x_i(t)$ are real for any $t$ and $x_{12}^{2}(t)\geq C>0$, Fig.~\ref{2p>>0}. The world lines 
of both particles run from minus to plus $t$-infinity and we have repulsion of particles in this case. But situation changes 
essentially if $C>x_{0,12}^{2}$. In this case $q_{12}$ and $p_{12}$ are pure imaginary, so real $x_i(t)$ exist in the 
finite interval of $t$ only, and in this interval $C\geq x_{12}^{2}(t)>0$, Fig.~\ref{2p>0}. Thus here we have creation of a 
pair of particles at some finite moment of time. These particles scatter with infinite velocities, slow down, stop and move 
to meet one another. At this moment they again reach infinite velocities and annihilate. 

If the induced system is given by (\ref{p2}) with $C<0$, we always have $x_{0,12}^{2}>0>C$, so by (\ref{p10}), (\ref{p11}) 
real solutions $x_i(t)$ exist outside of a finite interval of time, where $x_{12}^{2}(t)\geq0$. We have, Fig.~\ref{2p<0}, 
two particles coming from infinity: they attract, speed up and bump into each other (when the r.h.s.\ of (\ref{p11}) 
vanish). At this moment they reach infinite velocities and mutually annihilate. For a period induced system does  not exist: 
the r.h.s.\ of (\ref{p11}) is negative. Later, when again it reaches zero, two particles arise in some point of space with 
infinite velocities. Then they slow down and blow to infinity.
\begin{figure}[ht]
\begin{multicols}{2}
\hfill
\includegraphics[width=50mm]{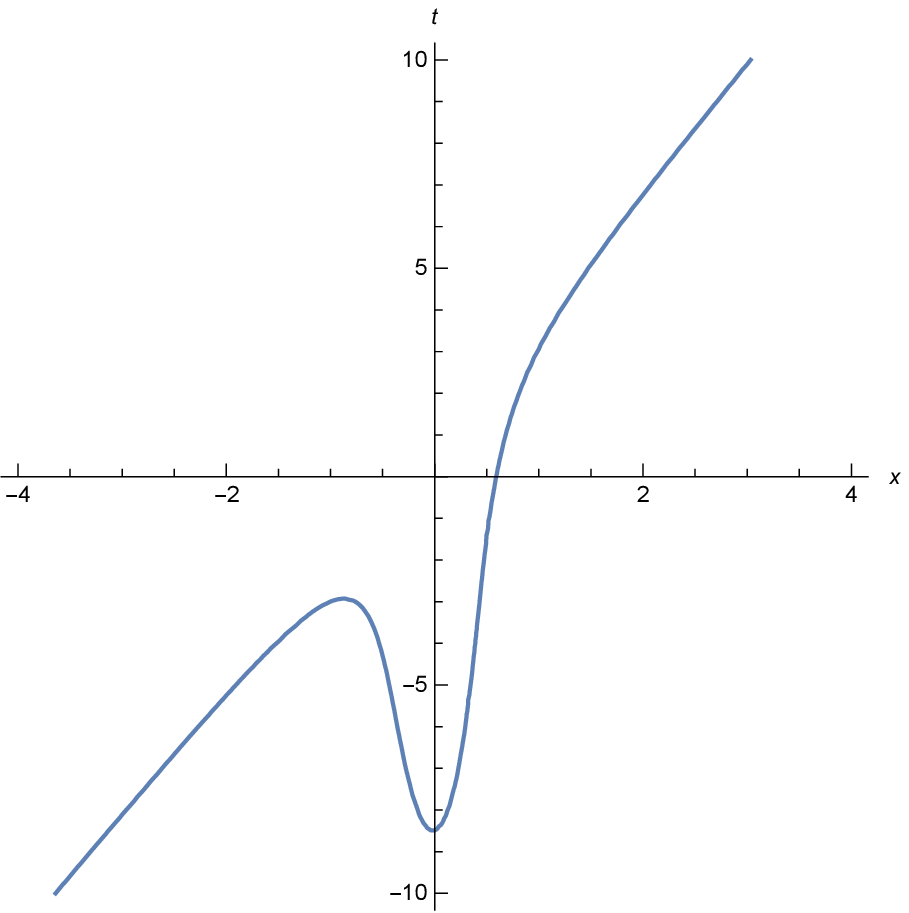}
\hfill
\caption{``Three'' particles}
\label{3-2p}
\hfill
\includegraphics[width=50mm]{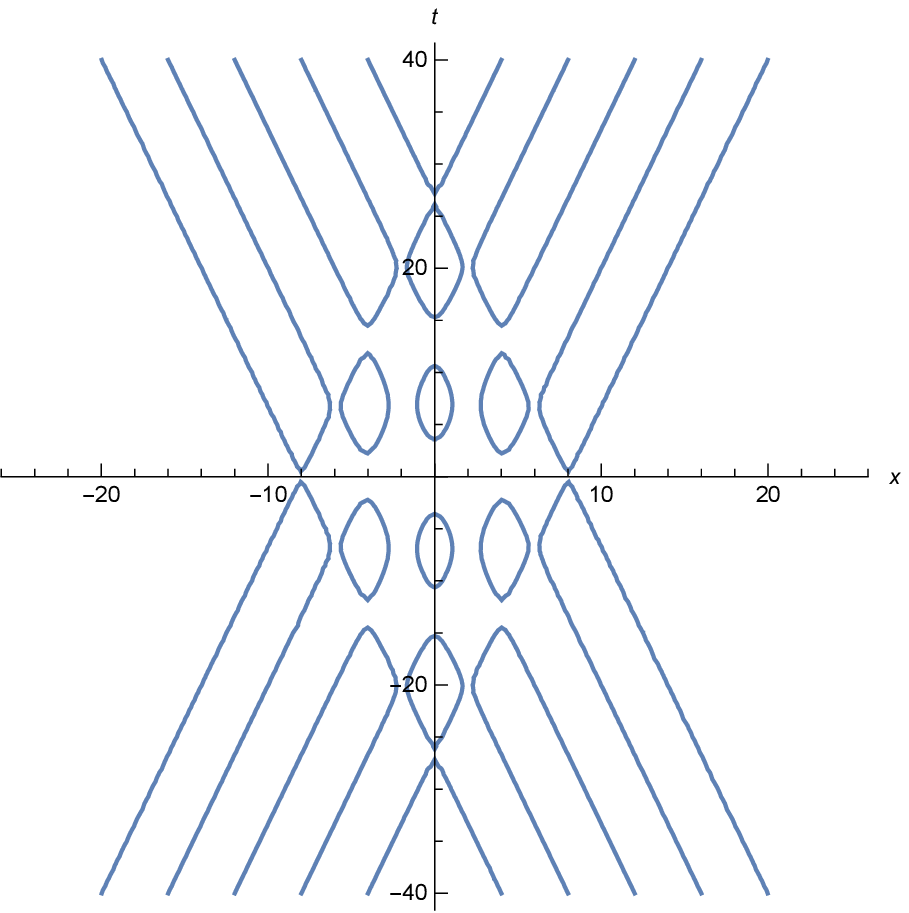}
\hfill
\caption{Ten particles}
\label{10p}
\end{multicols}
\end{figure}
So we have creation of particles that looks like a bang: particles appear from nowhere. Notice that variables of phase space 
$\mathcal{A}_2$ exist at any moment of time, independently of existence of the real solutions $x_i(t)$. In particular, all 
integrals of motion exist for any $t$. Following~\cite{LC} it is reasonable to use term ``cheshirization'' to denote such 
behavior of the induced system. It is necessary to emphasize that due to (\ref{p10}) and (\ref{p11}) singularities of the 
induced 
system---absence of real solutions---are movable: intervals where real solutions do not exist are defined by the initial 
data 
only. This property is analogous to the Painlev\'e property for integrable PDEs, while variables $q_i$ and $p_i$ are 
analogous to the scattering data for these equations. 

Analogously one can consider the higher values of $N$ in (\ref{p1}). Say, on the Fig.~\ref{3p} we present structure of 
the world lines of the system, given by $N=3$ in (\ref{p1}). By no means this structure is very unexpected for dynamical 
systems: three particles descend from infinity, two of them annihilate and for a period induced system has only one 
particle (only one real solution of the equation of the third order). Nevertheless, motion of this particle is far from 
being free: it slows down, stops and turns back. So it interacts with nonexistent particles of the induced system.  Another 
example 
of the case $N=3$, corresponding to another value of the constant $C$, is given on the Fig.~\ref{3-2p}. It is complimentary 
to the previous:  three particles exist in a finite interval of time only. Again we have creation and annihilation of 
particles and again dimension of configuration space of the induced system is variable. Despite the existence of only one 
particle asymptotically, we have system of three particles here: its solution is fixed by means of six initial data, 
positions and velocities of the particles at some moment. Because of translation invariance of the dynamical system, this 
moment always fits into interval where all three solutions do exist.

Finally, we present behavior of particles of the induced systems, Fig.~\ref{10p}, for the case $N=10$, i.e., for the 
counter-beams of five particles. We see that there could be many annihilation-creation effects and some cascades of 
intermediate (virtual) particles.

\subsection{Relativistic case.}\label{relat}
Examples of polynomial \textbf{relativistic induced system} are very close to nonrelativistic ones, considered above. Say, 
in the case $N=2$ the relativistic system in terms of the cone variables (\ref{a2'}) can be chosen as
\begin{equation}
 (\xi-q_1)(\xi-q_2)=\dfrac{C}{4}\biggl(\dfrac{1}{p_1^2}+\dfrac{1}{p_2^2}\biggr),\label{r1}
\end{equation}
where $q_{i}(\eta)=q_{0,i}-\eta/p_i^2$, cf.\ (\ref{a4}). It is easy to see that this equation is invariant with respect to 
transformation (\ref{a7}). Omitting details we write down equations of motion:
\begin{equation}
\xi''_{i}=\dfrac{(-1)^{i}C(\xi'_1-\xi'_2)^2(\xi'_1+\xi'_2)}{2(\xi_1-\xi_2)[(\xi_1-\xi_2)^{2}+C(\xi'_1+\xi'_2)]},
\quad i=1,2,\label{r2}
\end{equation}
where prime denotes derivative with respect to $\eta$. Taking that $\xi'_1+\xi'_2$ is integral of motion into account, we 
see that system (\ref{r2}) is close to the RS system (\ref{rs0}), cf.\ also (\ref{p6}) above. Behavior of the world lines on 
the $(x,t)$-plane is analogous to those given on Figs.~\ref{2p>>0}--\ref{3-2p}, while we have to mention that in cone 
variables condition on a world line to be time-like sounds as  $\xi'_i<0$. Here, instead of (\ref{p4}), the generic 
structure of the world lines is defined by the signs of the constant $C$ and of the l.h.s.\ of the equality
\begin{equation}
(\xi_1-\xi_2)^{2}+C(\xi'_1+\xi'_2)=q_{12}^{2}.\label{r3}
\end{equation}
Let the l.h.s.\ of (\ref{r3}) be positive, say, at $\eta=0$ and $C>0$. Then $q_{12}(\eta)$ by (\ref{a4}) is real for any 
$\eta$. Thus we have repulsion of two massive particles, close to that presented on Fig.~\ref{2p>>0}, with the only 
difference that particles moves with velocities less then the light one. If $C>0$ and the l.h.s.\ is negative at some 
initial moment of $\eta$, we have that particles exist in the finite interval of $\eta$ (or of $t$) only, i.e., their world 
lines behave as on the Fig.~\ref{2p>0}. It is absolutely unexpected that particles at the moments of creation and 
annihilation have infinite velocities, while the system (\ref{r3}) is relativistic. With growing of $\eta$ both particles 
slow down, their velocities become pre-light. After a period the particles accelerate, their velocities exceed velocity of 
the light and reach infinite values at point of annihilation. Finally, at the case of $C<0$ we have attraction, cf.\ 
Fig.~\ref{2p<0}, and real values of $\xi_1(\eta)$ and $\xi_2(\eta)$ exist outside a finite interval of $\eta$. In vicinity 
of collision particles behaves as above, and again we have an interval of cheshirization, after which particles appear again 
with infinite velocities and slow down to pre-light values. It is easy to see that in the relativistic case we have behavior 
of particles analogous to shown on Figs.~\ref{3p}--\ref{10p}, while asymptotically all word lines are time-like. 

\section{Concluding remarks}

\begin{figure}[ht]
\begin{multicols}{2}
\hfill
\includegraphics[width=50mm]{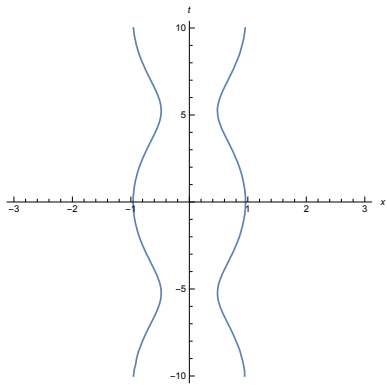}
\hfill
\caption{Two oscillating particles.}
\label{sinh5}
\hfill
\includegraphics[width=50mm]{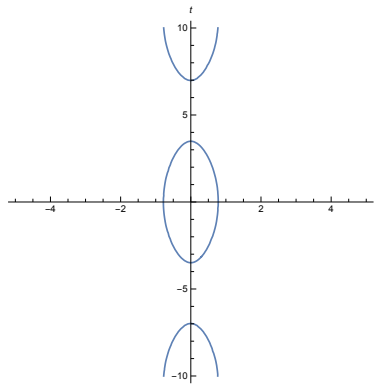}
\hfill
\caption{Cascade of virtual particles.}
\label{sinh2}
\end{multicols}
\end{figure}

In Sec.\ 4 we considered the simplest examples of the induced dynamical systems that describe, nevertheless, highly 
nontrivial collisions of particles. At the same time, examples based on the polynomial function $f$ in (\ref{i2}) do not 
lead to asymptotic phase shifts. Moreover, in case of polynomial $f$ dimensions of the real and complex solutions of 
(\ref{i2}) are equal, so in this sense polynomial examples are misleading, because condition that only real roots of 
(\ref{i2}) are considered looks artificial. On the other side in Sec.~2 it was shown that the condition of reality is 
essential for the induced system. Indeed, if function $f$ in (\ref{i2}) is given by (\ref{a20}), or (\ref{a5}), we get 
infinitely many roots in the complex domain of $x$. These roots cannot correspond to a dynamical system, because (real) 
dimension of the space $\mathcal{A}_N$ is $2N$. This is confirmed also by the following example that generalizes example in 
Sec.~4.1 and demonstrates new interesting features of the induced systems. 

Let us consider $f(\mathbf{q},\mathbf{p})=\sinh{q}_1\sinh{q}_2-C$ instead of (\ref{p2}), were again $C$ is a real constant, 
so that 
(\ref{i2}) takes the form $\sinh({q}_1-x)\sinh({q}_2-x)=C$. Equations on sum $x_1(t)+x_2(t)$ are exactly as the first 
equations in (\ref{p4})--(\ref{p6}), while for the differences (\ref{p3}) we have now equality 
$\cosh{x}_{12}=\cosh{q}_{12}+2C$, cf.\ (\ref{p4}). Correspondingly, equation of motion of the induced system in this case 
takes the form
\begin{equation}
 \ddot{x}_j=\dfrac{(-1)^{j}C{\dot{x}_{12}}^{2}[\cosh^{2}x_{12}-
 2C\cosh x_{12}+1]}{\sinh x_{12}[(\cosh x_{12}-2C)^{2}-1]},\quad j=1,2.\label{c1}
\end{equation}
We have here two systems in correspondence to the sign of the constant $C$, like in Sec.~4.1. If $C>0$ and initial value
obeys $\cosh x_{12}(0)>1+2C$, then for any $t$ there exist two repulsing particles which world lines are close to those on 
Fig.~\ref{2p>>0}. But if $1>\cosh x_{12}(0)-2C>-1$, we have to consider two cases: $C>1$ and $1>C>0$. In the 
first case we get two particles that oscillate at around their central positions, see Fig.~\ref{sinh5}, while distance 
between their centers being fixed. In the second case we get behavior shown on the Fig.~\ref{sinh2}, that can be interpreted 
as cascade of virtual particles. In this case we have periodic sequence of the creations/annihilations of particles. In the 
case $C<0$ real solutions exist outside some finite interval of time only, we have again effect of cheshirization and 
behavior of the world lines is close to the one on Fig.~\ref{2p<0}. In all these cases assuming complex values of $x_j$ we 
get a strange nets of roots, $x_j\to x_j+2\pi ik_j$, $k_j\in\mathbb{Z}$, that do not admit any dynamical interpretation 
from our point of view.

Some examples of induced systems considered above admit explicit derivation of differential equations of motion of the 
Newton type. In this respect an interesting problem is existence of matrices $W$, besides those given in (\ref{cm9}), that 
generates by (\ref{cm7}) explicit dynamical systems different from CM and RS ones. On the other side, examples 
from Sec.~4 show that cases, where forces in the r.h.s.\ of (\ref{i7}) are explicit, are very rare and in generic situation 
investigation of induced systems must be based on the description of solutions of Eq.~(\ref{i2}). 

Unexpected properties of the induced dynamical systems of the kind (\ref{p6}), (\ref{r2}) and (\ref{c1}), involving such 
``quantum'' effects as creation/annihilation of particles, bound states, virtual particles, etc., are new to our knowledge 
and deserves further investigation. It is also interesting to generalize consideration of the Sec.~2---dynamics of 
singularities of the nonlinear equations---to the case where solitons are unstable and, say, collision of two regular 
solitons leads to singularity. Such soliton solutions were studied in \cite{orl}, \cite{FST}, and \cite{BZ}, in particularly 
for the case of the Boussinesq equation.

\noindent\textbf{Acknowledgment.} Author thanks L.~V.~Bogdanov, A.~M.~Liashyk and A.~V.~Zotov for fruitful discussions.


\begin{thebibliography}{99}
\frenchspacing
\bibitem{kdv} Arkad'ev V. A., A. K. Pogrebkov, Polivanov M. K., ``Singular solutions of the KdV equation and the 
method of the inverse problem,'' \textsl{Zap. Nauchn. Sem. LOMI} \textbf{133} (1984) 17--37.
\bibitem{sinh1} A. K. Pogrebkov, ``Singular solitons: an example of a sinh-Gordon equation,'' \textsl{Lett. Math. Phys.} 
\textbf{5}:4 (1981) 277--285. 
\bibitem{sinh2} A. K. Pogrebkov, Polivanov M. K., ``Interaction of particles and fields in classical theory,'' 
\textsl{Soviet J. Particles and Nuclei} \textbf{14}:5 (1983) 450--457. 
\bibitem{sinh3} A. K. Pogrebkov, Polivanov M. K., ``The Liouville and sinh-Gordon equations. Singular solutions, dynamics of 
singularities and the inverse problem method,'' in \textsl{Mathematical physics reviews} Vol. 5, \textsl{Soviet Sci. Rev. 
Sect. C Math. Phys. Rev.}, Harwood Academic Publ., (1985) pp. 197--271.
\bibitem{nov}  Novikov S., Manakov S.V., Pitaevskii L.P., Zakharov, V.E., \textsl{Theory of Solitons: The Inverse Scattering 
Method} Springer (1984), 276pp.
\bibitem{CM1} F. Calogero, ``Exactly solvable one-dimensional many-body problems'', \textsl{Lett. Nuovo Cim.} \textbf{13}  
(1975) 411--416.
\bibitem{CM2} J. Moser, ``Three integrable Hamiltonian systems connected with isospectral deformations'', \textsl{Adv. 
Math.} \textbf{16}  (1975) 197--220.
\bibitem{RS1} S. N. M. Ruijsenaars and H. Schneider, ``A new class of integrable systems and its relation to solitons,'' 
\textsl{Annals of Physics (NY)} \textbf{170}(2)  (1986) 370--405.
\bibitem{RS2} S. N. M. Ruijsenaars, ``Complete integrability of relativistic Calogero-Moser systems and elliptic function 
identities,'' \textsl{Commun. Math. Phys.} \textbf{110}:2 (1987) 191--213.
\bibitem{OP} M.A. Olshanetsky, and A.M. Perelomov,``Explicit solution of the Calogero model in the classical case and 
geodesic flows on symmetric spaces of zero curvature,'' \textsl{Lett. Nuovo Cimento (2)} \textbf{16}:11 (1976) 
333--339.
\bibitem{RS3}  S.N.M. Ruijsenaars, ``Action-angle maps and scattering theory for some finite-dimensional integrable systems. 
I: The pure soliton case,'' \textsl{Commun. Math. Phys.} \textbf{115} (1988) 127--165.
\bibitem{Kr} I.M. Krichever, ``Elliptic solutions of the Kadomtsev--Petviashvili equation and integrable systems of 
particles,'' \textsl{Funct. Anal. Appl.} \textbf{14}  (1980) 282--290.
\bibitem{GP} L. Gavrilov, and A.M. Perelomov, ``On the explicit solutions of the elliptic Calogero system,'' \textsl{J. 
Math. Phys.} \textbf{40}  (1999) 6339--6352.
\bibitem{LC} Lewis Carroll, \textsl{Alice's Adventures in Wonderland} (1865)
\bibitem{orl} A. Yu. Orlov, ``Collapse of solitons in integrable model,'' \textsl{Preprint IAIE} No. 221 (1983) IAIE, 
Novosibirsk.
\bibitem{FST} G. E. Falkovich, M. D. Spector, S. K. Turitsyn, ``Destruction of stationary solitons and collapse in the 
nonlinear string equation,'' \textsl{Phys. Lett. A} \textbf{99} (6--7) (1983) 271--274.
\bibitem{BZ} L. V. Bogdanov, V.E. Zakharov, ``The Boussinesq equation revisited,'' \textsl{Physica D} \textbf{165} (2002) 
137--162.
\end{thebibliography}
\end{document}